\documentclass[namedreferences]{solarphysics}
\usepackage[optionalrh,natbib]{spr-sola-addons} 
\usepackage{graphicx}        

\usepackage[pdfborder={0 0 0 },urlcolor=blue,breaklinks]{hyperref}

\usepackage{graphicx}
\usepackage[usenames]{color}

\usepackage{subeqn}

\newcommand{\eal}{{\itshape{et al.}}}
\def\p{\partial}
\def\Om{{\it \Omega}}

\newcommand{\BoldVec}[1]{\mathchoice%
  {\mbox{\boldmath $\displaystyle     #1$}}%
  {\mbox{\boldmath $\textstyle        #1$}}%
  {\mbox{\boldmath $\scriptstyle      #1$}}%
  {\mbox{\boldmath $\scriptscriptstyle#1$}}%
}
\newcommand{\EQ}{\begin{equation}}
\newcommand{\EN}{\end{equation}}
\newcommand{\EQA}{\begin{eqnarray}}
\newcommand{\ENA}{\end{eqnarray}}

\newcommand{\uu}{\mathbf{u} {}}

\newcommand{\vv}{\mathbf{v} {}}

\newcommand{\UU}{\mathbf{U} {}}

\newcommand{\bb}{\mathbf{b} {}}

\newcommand{\BB}{\mathbf{B} {}}

\newcommand{\ee}{\mathbf{e} {}}

\newcommand{\kk}{\mathbf{k} {}}

\newcommand{\nab}{\BoldVec{\nabla} {}}

{}

\def\etat{\eta_{\rm t}}
\def\etaT{\eta_{\rm T}}

\newcommand{\cm}{\,{\rm cm}}

\long\def\symbolfootnote[#1]#2{\begingroup%
\def\thefootnote{\fnsymbol{footnote}}\footnote[#1]{#2}\endgroup}

\begin{document}

\begin{article}

\begin{opening}

\title{Solar Dynamo and Toroidal Field Instabilities\symbolfootnote[1]{Invited talk}}

\author{Alfio \surname{Bonanno}}
\institute{INAF, Osservatorio Astrofisico di Catania, Via S.Sofia 78, 95123 Catania, Italy  \email{alfio.bonanno@inaf.it}} 

\runningauthor{A. Bonanno}
\runningtitle{Solar Dynamo and MHD instabilities}

\date{}

\begin{abstract}
The possibility of non-axisimmetric (kink) instabilities of a toroidal field seated in the tachocline is much discussed in the literature.
In this work, the basic properties of kink and quasi-interchange instabilities, 
produced by mixed toroidal and poloidal configuration, will be briefly reviewed. In particular it will be shown 
that the unstable modes are strongly localized near the Equator and not 
near the Poles as often claimed in the literature.  Based on the results of recent numerical simulations, 
it is argued that a non-zero helicity can already be produced at a non-linear level.
A mean-field solar dynamo is then constructed with  a {\it positive} $\alpha$-effect in the overshoot layer
localized near the Equator and a meridional circulation with a deep return flow.
Finally, the possibility that the solar cycle is driven by  a $\alpha\Omega$ dynamo generated
by the negative subsurface shear in the supergranulation layer will also be discussed. 
\end{abstract}

\end{opening}
\section{Introduction}
\label{S-Introduction}
An essential ingredient of any  dynamo model is the $\alpha$-effect,  where $\alpha$ is the
transport coefficient of the closure relation for the turbulent electromagnetic force. Owing  to its pseudoscalar nature this term represents
a likely possibility to produce a dynamo action with an axially symmetric field. 

From the theoretical point of view it is expected that both the kinetic helicity $\langle \vv \cdot \nabla \times  \vv \rangle$
and the current helicity $\langle \bb \cdot \nab \times \bb \rangle$ produced by the velocity  $\vv$ and magnetic field $\bb$ fluctuations 
contribute to this term \citep{gruda}, although the relative importance of these terms is still a subject of debate \citep{sila}. 
On physical grounds it is conceivable that in the bulk of the convection zone
 ``cyclonic" turbulence  \citep{Par55} could be an efficient mechanism to produce a turbulent dynamo whose $\alpha$-effect is dominated by the kinetic helicity \citep{skr66}. 

The discovery, due to the interpretation of helioseismic data  \citep{schou},  that the stable stratified region 
below the convection zone is characterized by the presence of a  strong horizontal 
shear \citep{kos96} has suggested the possibility of an $\alpha$-effect located just beneath the convection zone  \citep{Par93} 
in the region called the tachocline  \citep{SZ92}.  
This fact has opened the door to the possibility that the source of the $\alpha$-effect can have a magnetic origin, 
which can thus be attributed to various possible MHD instabilities \citep{fema,dikpati01,pijab}.
More recently,  \cite{bu11,bu12} have further clarified a few aspects of kink and quasi-interchange
instabilities in stably stratified plasma. While kink instabilities are generated from a pure toroidal field, 
quasi-interchange instabilities are produced by a mixed combination of the poloidal and toroidal field
and their spectrum can be rather different from simple kink waves. It has also been shown that kink modes in spherical 
geometry are more effective near the Equator, a result that support the possibility of 
producing  a non-zero $\alpha$-effect at low latitudes. 
In fact, although the presence of a poloidal field breaks the symmetry creates a preferred helicity in the turbulence flow, 
\cite{fabio} recently argued that kink instabilities alone can produce a preferred chirality in the turbulent plasma due to a symmetry-breaking effect 
at non-linear level. 

In this article, after reviewing the basic properties of kink and quasi-kink instabilities in a stably stratified plasma, 
the possibility of explicitly constructing a solar dynamo with an  $\alpha$-effect localized at the bottom of the convective
zone and at low latitudes will be discussed and compared to mean-field models where the $\alpha$-effect is instead localized
near the surface layer, as proposed by \cite{axel05}.

\section{Quasi-Interchange Instabilities in a Stably Stratified Plasma}
The stability of a stably stratified column of plasma in the ideal MHD limit is the central problem of most of the controlled-fusion
literature.  In this context, the energy principle \citep{bernstein}  has extensively been used in the past to study the stability of 
poloidal or toroidal fields \citep{tay73a,tay73b} and also of mixed combinations of both \citep{tay80}. 

In cylindrical geometry, it can be 
proven that the plasma is stable for all azimuthal and vertical wave numbers ($m$ and $k$), if it is stable for $m=0$ in the $k\rightarrow 0$ limit,
and for $m=1$ for all $k$ \citep{gopo}.  On the other hand, to show that a generic configuration with a 
combination of axial field and non-homogenous azimuthal field is stable against the $m=1$ mode (for all $k$) is not an easy task in general and one has to resort  either to a variational approach or to a numerical investigation of the full eigenvalue problem in the complex plane. 
In this respect, the ``normal mode" approach can be more useful in astrophysics, as 
it is often important to know the growth rate of the instability and the properties of the 
spectrum of the unstable modes \citep{bo08a,bo08b}. 
In particular, generic combinations
of axial and azimuthal fields are subject to a class of resonant MHD 
waves that can never be stabilized for any value of the 
ratio of poloidal and toroidal fields.  The instability of these waves has a mixed 
character, being both current- and pressure-driven. In 
this case the most rapidly growing unstable modes are resonant, {\it i.e.} the wave vector 
$\kk = (m/s) {\mathbf{e}}_\theta + k_z \mathbf{e}_z $ is perpendicular to the magnetic 
field, $\BB \cdot \kk = 0$ where $k_z$ is the wavevector in the axial 
direction, $m$ is the azimuthal wavenumber, and $s$ is  the cylindrical radius.
The length scale of this instability depends on the ratio of poloidal and 
azimuthal field components and it can be very short, while the width of the 
resonance turns out to be extremely narrow. For this reason its excitation in  
simulations  can be problematic. 

It is interesting to have a qualitative understanding of the MHD spectrum for a simple cylindrical plasma equilibrium configuration 
consisting of a mixed configuration of an azimuthal $B_\phi$ and a constant axial field $B_z$. 
As was shown in \cite{bu11}, for a generic disturbance of the form $\mathrm{e}^{(\sigma t - \mathrm{i} k_z z - \mathrm{i} m \varphi)}$
an approximate expression for the dimensionless growth rate
$[\Gamma=\sigma/\omega_{A\phi}]$ being  $\omega_{A\phi}$   the  Alfv\'en frequency in the azimuthal direction, is given by
\begin{equation}
\Gamma^2=\frac{2 m^2 (\alpha-1)}{m^2+(p^2+m^2) \varepsilon^2},
\label{sette}
\end{equation}
where $p$ is the dimensionless radial wavenumber, $\varepsilon=B_z/B_\varphi$,  $\alpha=\partial \ln B_\varphi /\partial \ln s$, and $s$ is the cylindrical radius.
Moreover the dimensionless vertical wavenumber $[q=k_z s]$ is close to the resonance condition [$f\equiv q \varepsilon +m \approx 0$] which implies
that the total Alfv\'en frequency is zero $[\omega_A = \BB \cdot \kk /\sqrt{4\pi \rho}=0]$ at the resonance. The instability is never suppressed for any finite value of $\varepsilon$ and 
the growth rate is a rapidly increasing function of $m$ in particular
 $\Gamma^2 \approx 
(1+\varepsilon^2)^{-1}$ in the limit $m \gg p^2$.  
If $\alpha < 1$, it is possible to show that 
\begin{equation}\label{ggr}
\Gamma^2 \approx f^2 \frac{1 + \alpha}{1 - \alpha},
\end{equation} 
that implies instability  if $\alpha > -1$. 
The profile with 
$\alpha < -1$ is stable in this approximation. Note that modes with $q$
satisfying the resonance condition $\omega_A = 0$ (or $f=0$) are marginally stable 
because $\Gamma=0$ for them, but $\Gamma^2 > 0$ in a neighborhood of the 
resonance. Therefore, the dependence of $\Gamma$ on $q$ should have a two-peak 
structure for any $m$. As in the case $\alpha > 1$, the instability occurs for any value of  
$\varepsilon$. If $\alpha =1$, then we have
\begin{equation}
\Gamma^2 \approx \mu f \left[ \frac{2m}{m^2 + q^2} \pm \sqrt{ \frac{4 m^2}{(m^2 + q^2)^2}
+ 4 \mu} \right].
\end{equation}
where $\mu$ is a positive number of the order unity.
In this case, the dependence $\Gamma^2(q)$ also has a two-peak structure because 
$\Gamma = 0$ at the resonance but $\Gamma^2 > 0$ in its neighborhood. The 
instability is always present for any finite value of $\varepsilon$.
This explicit solution shows that, if $\alpha > -1$, the instability always occurs for 
disturbances with $q$ and $m$ close to the condition of magnetic resonance, 
$\omega_A =0$. The axial field cannot suppress the instability which occurs 
even if $B_z$ is significantly greater than   $B_{\varphi}$. 

\section{Kink Instabilities below the Tachocline}
Can the instabilities described in the previous section be operative in the overshoot layer of the Sun, or even below?
To address this question we must consider the problem in spherical geometry, including the stabilizing effect
of gravity and the destabilizing effect of thermal diffusivity. In fact the 
stability of the spherical magnetic configurations has been studied in much less 
detail and even the overall  stability properties of stellar radiation zones are rather unclear.
\cite{brano06} studied 
the  stability of a random initial field in the stellar radiative zone  by direct numerical simulations,
and it was found that the stable magnetic configurations generally have the form of tori with 
comparable poloidal and toroidal field strengths. The possible relaxation mechanism was further discussed by 
\citet{duma10}. The stability of azimuthal fields near the rotation axis has also been studied by 
\cite{spru99}. The author used a heuristic approach to estimate the growth rate and criteria of instability.
Unfortunately, many of these estimates and criteria are misleading because they do not apply in the main
fraction of the volume of a radiation zone where the stability properties can be qualitatively different.
The heuristic approach was criticized by \cite{za07}. The stability of the toroidal 
field in rotating stars has been considered by \cite{kit08} and 
\cite{kiru08} who argued that the magnetic instability is 
essentially three-dimensional and determined the threshold field strength 
at which the instability sets. Estimating this threshold in the solar 
radiation zone, the authors impose an upper limit on the magnetic field  
$\approx 600$ G.

The problem has recently been investigated by \cite{bu12} where it was shown that the most unstable 
modes have low radial wavelengths at variance with the claim of \cite{kiru08}. Moreover 
if the thermal conductivity is considered, no threshold field is needed to trigger the instability.
The most interesting result is illustrated by the angular dependence of the growth rate, as can be seen in Figure 1 where 
it appears that the instability is effective mostly near the Equator, and not along the axis, as originally supposed by \cite{tay73a}.
\begin{figure}
\begin{center}
\includegraphics[width=10cm]{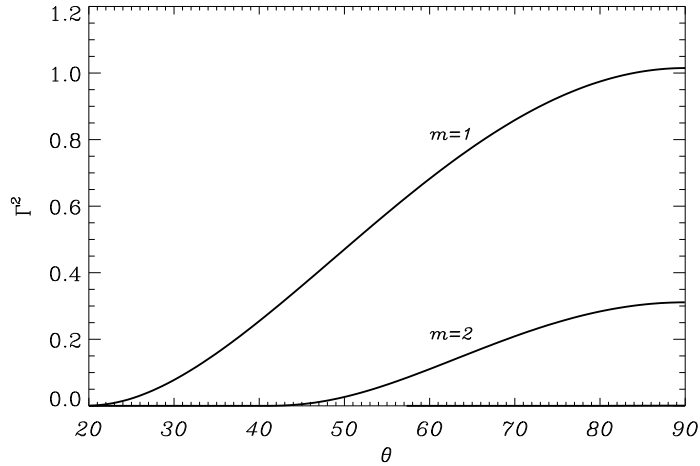}\label{lar}
\caption{The dimensionless growth rate of the fundamental radial mode 
as a function of the polar angle, $\theta$, for $\ell=5$, $m=1,2$,  
and $d=0.1$ 
in the case of a neutral stratification.
Note that $\ell$ is the latitudinal ``quantum" number of the perturbation. See \cite{bu12} for further details.} 
\end{center}
\label{fig:1uno}
\end{figure}
{In general the location of the most unstable latitude depends also on the geometry of the basic state and cannot be
determined only by local instability criteria.}

Can this instability produce an $\alpha$-effect? 
{It is difficult to  consistently compute the strength of the $\alpha$-effect 
within the linear analysis alone, although a first estimation in this direction has been 
provided by \cite{ru11}}. However, recent investigations in cylindrical symmetry pointed out that it is possible to generate a non-zero helicity out of a kink instability due  to non-linear effects \citep{fabio,sysb}.

\section{Flux-Transport Dynamo and Tachocline $\alpha$-Effect}
The previous discussion put forward that a likely location of the $\alpha$-effect  generated by current-driven instabilities
is precisely in the overshoot layer with a strong concentration of the turbulence at low latitudes
where the instability is more effective. 
Although only 3D numerical simulations can in principle determine the full structure of the $\alpha$-tensor, 
as a first step in building a realistic model we can consider 
a latitudinal dependence of the type $\cos \theta \sin^2\theta$ to model the suppression of the $\alpha$-effect near the poles.
The other essential ingredient is the inclusion of the meridional circulation. 
In fact, in  the presence of a low eddy diffusivity  [$\etat$] (as is likely to be in the overshoot region),  
the magnetic Reynolds number becomes very large and the dynamics of the mean-field flow becomes 
an essential ingredient of the dynamo process. In this regime
the advection produced by the meridional circulation dominates the diffusion of the magnetic field which
is then ÒtransportedÓ by the meridional circulation \citep{dikpati99,dikpati01,manf01,bo02,pij04,bo06,gue08}.

The magnetic induction equation reads
\EQ\label{induction}
{\partial {\BB} \over \partial t} = {\nab \times} ({\UU}
\times {\BB} + \alpha {\BB}) - {\nab} \times \left( \etaT {\nab} \times {\BB}\right),  
\EN
where $\etaT$ is the turbulent diffusivity.  Axisymmetry  implies that relative to spherical coordinates the magnetic field (${\BB}$) 
and the mean flow field (${\UU}$), respectively, read
\[
\BB = B(r,\theta,t){\ee}_\phi+ {\nab}\times [ A(r,\theta,t) \ee_\phi]
\]
\[
\UU = {\uu}(r,\theta) + r\sin\theta \Omega(r,\theta)\ee_\phi\nonumber
\]
being $ A(r,\theta,t)$ the vector potential.
The meridional circulation [${\uu}(r,\theta)$]
and differential rotation [$\Omega(r,\theta)$] are the poloidal and toroidal components
of the global velocity flow field $[{\UU}]$.
In particular the poloidal and toroidal components of Equation (\ref{induction}) respectively determine 
\begin{subequations}
\label{ddd-coupled}
\begin{eqnarray}
&&{\partial A \over \partial t}
+\frac{1}{s}({\uu}   {\cdot} {\nab}) (sA)  = \alpha B +\frac{{\etaT}}{r}\frac{\p^2 (rA)}{\p r^2}
+\frac{\etaT}{r^2}\frac{\p }{\p \theta}\Bigl( \frac{1}{s}\frac{\p (sA)}{\p \theta}\Bigr), \\[2mm]
&&\frac{\p B}{\p t} + s\rho({\uu} {\bf \cdot}  {\nab}) \Bigl(\frac{B}{s\rho}\Bigr)=
\frac{\p \Om }{\p r}\frac{\p (A\sin\theta)}{\p \theta}
-\frac{1}{r}\frac{\p \Omega}{\p \theta}
\frac{\partial (sA)}{\partial r}+\frac{1}{r} \frac{\p}{\p r}\Big({\etaT}\frac{\p (rB)}{\p r} \Big)\nonumber\\
&&+\frac{\etaT}{r^2}\frac{\p }{\p \theta}\Bigl(\frac{1}{s}\frac{\p (sB)}{\p \theta}\Bigr)
-\frac{1}{r}\frac{\p}{\p r} \Big (\alpha \frac{\p (rA)}{\p r} \Big)
-\frac{\p}{\p\theta}\Bigl(\frac{\alpha}{\sin\theta}\frac{\p (A\sin\theta)}{\p\theta}\Bigr),
\end{eqnarray}
\end{subequations}
%
where $s=r\sin\theta$. The $\alpha$-effect and the turbulent diffusivity are parametrized by means of 
\begin{eqnarray}
&&\alpha = \frac{1}{4}\, \alpha_0 \cos\theta \sin^2 \theta \Big [1+{\rm erf}\Bigl(\frac{x-a_{1}}{d}\Bigr)\Big ]\Big [1-{\rm erf} \Bigl(\frac{x-a_{2}}{d}\Bigr)\Big ],\nonumber\\
&&\eta = \eta_c+\frac{1}{2}(\eta_t-\eta_c)\Bigl[1+{\rm erf}\Bigl(\frac{r-r_\eta}{d_\eta}\Bigr)\Bigr],
\label{aa}
\end{eqnarray}
where $\alpha_0$ is the amplitude of the $\alpha$-effect, $x=r/R_\odot$ is the fractional radius, 
$a_1$, $a_2$ and $d$ define the location and the thickness of the 
turbulent layer, $\eta_t$ is the eddy diffusivity,  $\eta_c$ the magnetic diffusivity beneath the
convection zone and $d_\eta$ represents the width of this transition.
In this investigation, the  values $a_1=0.67$, $a_2=0.72$, and $d_\eta=0.025$ have been used.

\begin{figure}
\centering
\includegraphics[width=0.45\textwidth]{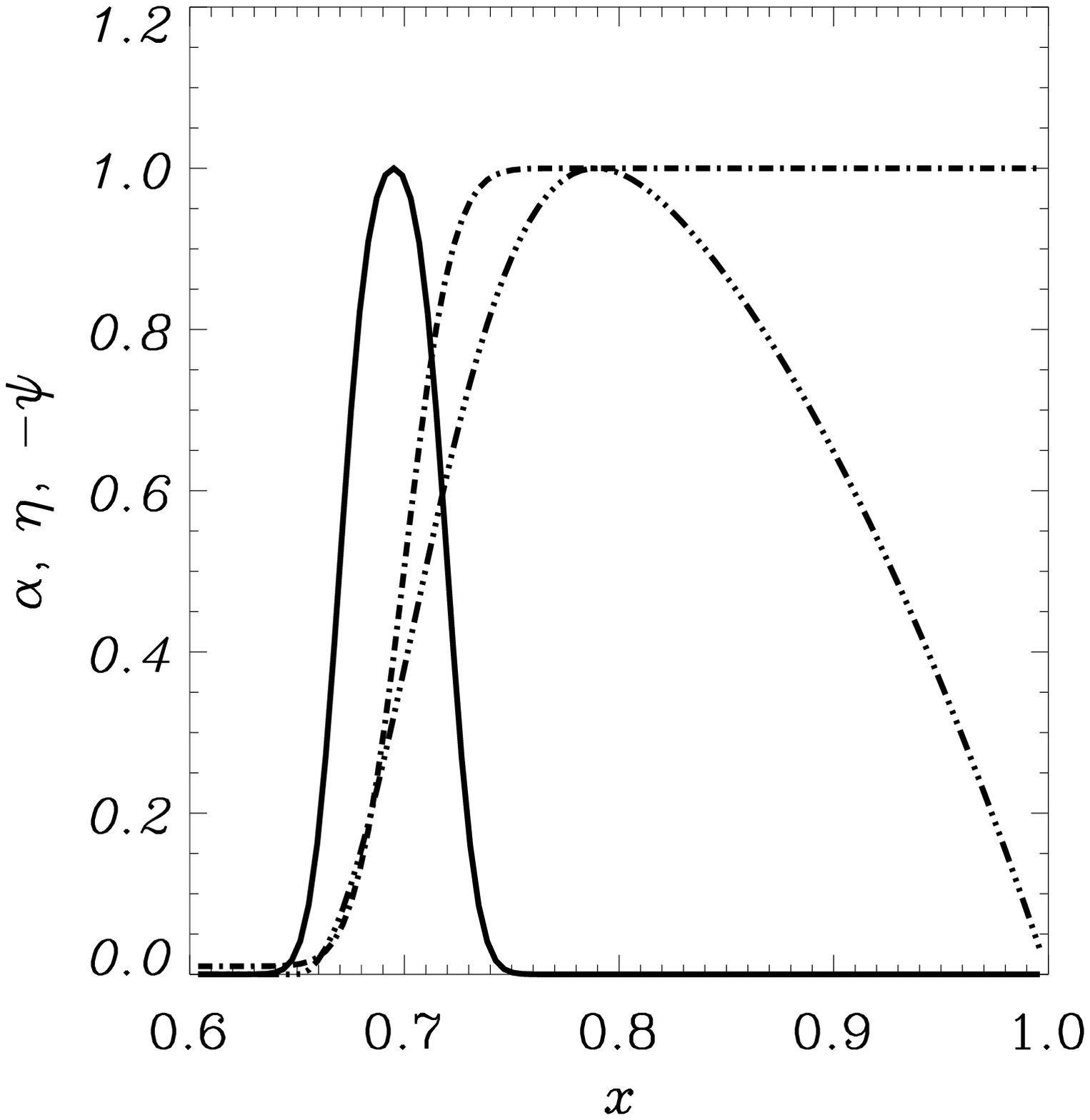}
\includegraphics[width=0.45\textwidth]{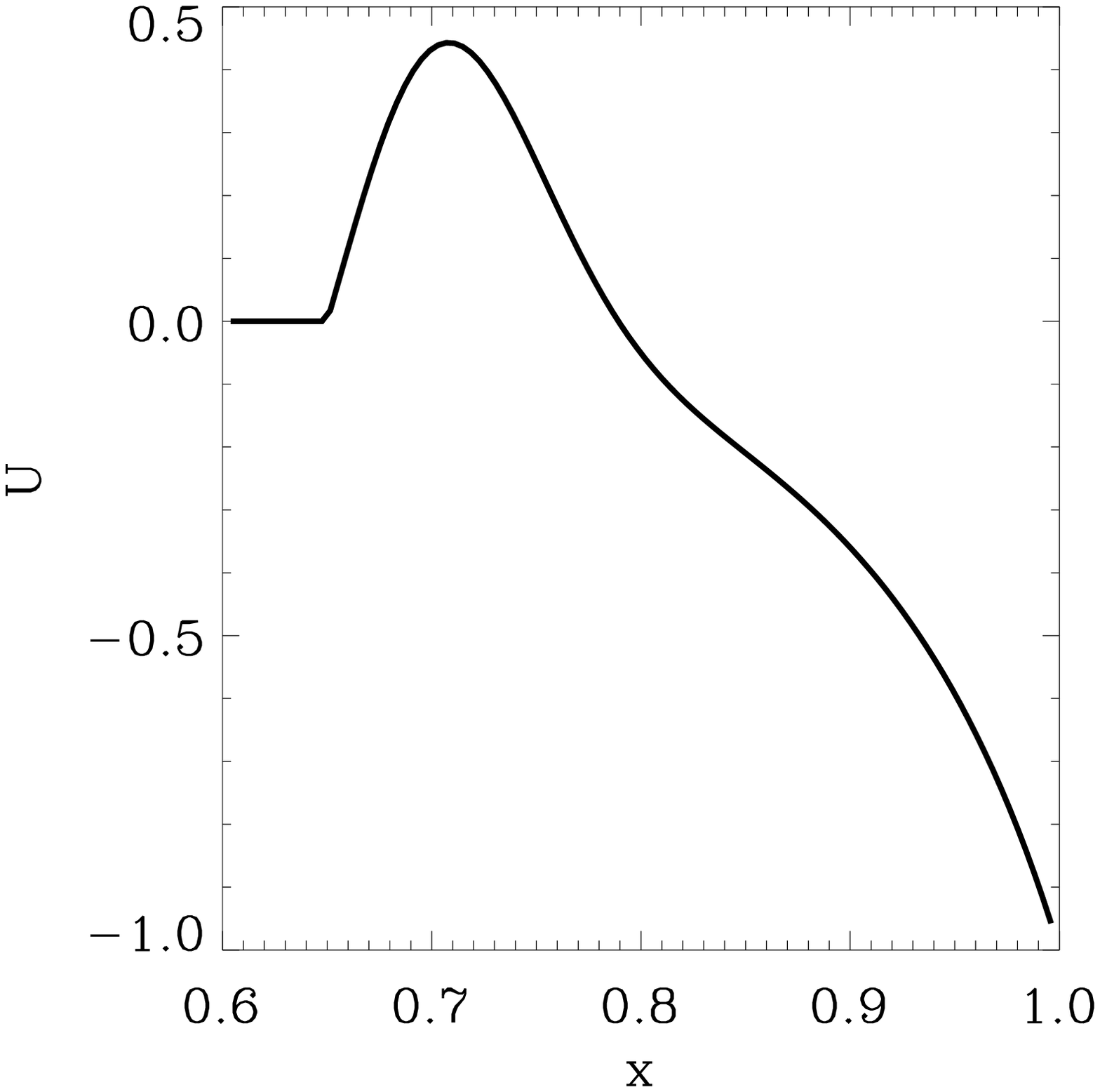}
\caption{The $\alpha$-effect (solid line), turbulent diffusivity (dot-dashed line) and (minus) the function $\psi(r)$ (dot-dot-dot dashed line), 
are depicted in the left panel. 
The meridional circulation in units of the maximum surface value
at a latitude of $45^\circ$ is instead depicted in the right panel.}
\label{fig:stream}
\end{figure}

The components of the meridional circulation can be represented with the help of a stream function 
$\Psi(r,\theta)=-\sin^2\theta\cos\theta\,\psi(r)$ 
so that 
\EQ
u_r= \frac{1}{r^2\rho\sin\theta} \frac{\partial \Psi}{\partial\theta},\;\;\;\;
u_\theta= -\frac{1}{r\rho\sin\theta}\frac{\partial \Psi}{\partial r}
\label{13}
\EN
with the consequence that the condition $ {\nab} {\cdot} (\rho {\uu}) = 0$ is automatically fulfilled. 
A strategy to constrain several properties of the 
meridional circulation is to assume the differential rotation profile 
$\Omega(r,\theta)$ as a given ingredient, and deduce 
an approximation for the function $\psi$ 
from the angular-momentum conservation along the azimuthal direction. An approximate expression for $\psi$  is thus 
\EQ\label{dupsi}
\psi\approx \frac{5\rho r  }{2 \Omega_{\rm eq} } \int_0^\pi \langle u_r u_\theta \rangle {\mathrm d}\theta
\EN
where  $\Omega_{\mathrm{eq}}$ is the equatorial angular velocity.
In particular, for the standard, isotropic mixing-length theory, Equation (\ref{dupsi}) becomes \citep{durney00}
\EQ\label{dupsi2}
\psi\approx - \frac{5\rho r  }{2 \Omega_{\rm eq} } \langle u_r^2\rangle\,.
\EN
In principle it would be possibile to explicitly compute $\psi$ and $\psi'$ using the relation 
(\ref{dupsi2}) knowing the convective velocities of the underlying stellar model.
In practice this would be problematic, because the convective fluxes and their radial derivatives computed from standard mixing-length theory are  discontinuous at the base of the convective zone. In a more realistic situation  the presence of an overshoot layer implies that
$\langle u_r^2\rangle \rightarrow 0$ smoothly
so that $u_\theta $ is continuous at the inner boundary. 
Nevertheless one can use the representation (\ref{dupsi2}) to determine the stagnation point where $\psi'=0$, which turns out to be
around $x=0.8$ in terms of the fractional radius.
An explicit form of the  function $\psi$ which incorporates the following features reads
\EQ
\psi=C \; \left[1-\exp{\left(-\frac{(x-x_b)^2}{\sigma^2}\right)} \right] (x-1) \; x^2 \,,
\EN
where $C$ is a normalization factor, $x_b=0.65$ defines the  penetration of the flow, $\sigma=0.08$ measures how rapidly
$\langle u_r^2\rangle$ decays to zero in the overshoot layer and the location of the stagnation point.
The density profile is taken to be
$\rho=\rho_0 \Bigl(\frac{1}{x}-x_0\Bigr)^m$
in which $m$ is an index representing the  the stratification of the underlying solar model, its value in the region of interest is approximately $1.5$, 
and $x_0=0.85$, so that with these value the strength of the meridional circulation at low latitudes is of the same order as the surface flow, 
as discussed by \cite{bo12}; see also \cite{pipin11} for a similar investigation.
The radial profile of the $\alpha$-effect, turbulent diffusivity, stream function and meridional circulation used in the 
calculation are depicted in Figure \ref{fig:stream}.

The  helioseismic profile for the differential rotation is taken so that 
\begin{equation}
\Omega(r,\theta)=\Omega_{\mathrm{c}}
+   \delta \left (\frac{r-r_c}{d_c}\right)
\bigl(\Omega_{\mathrm{s}}(\theta)-\Omega_{\mathrm{c}}\bigr), \quad 
\label{eq3}
\end{equation}
\noindent where $\Omega_{\mathrm{c}}/2 \pi=432.8$ nHz is the uniform angular
velocity of the radiative core,
$\Omega_{\mathrm{s}}(\theta)=\Omega_{\mathrm{eq}}+a_2\cos^2  \theta+a_4\cos^4 \theta$
is the latitudinal differential rotation at
the surface and $\delta(x)\equiv (1+{\rm erf}(x))/2$.
In particular $\Omega_{\mathrm{eq}}/2\pi$$=$$460.7$ nHz is the angular
velocity at the Equator,  $a_2/2\pi$$=$$-62.9$ nHz,  and
$a_4/2\pi$$=$$-67.13$ nHz.
In this calculation the angular velocity is normalized in terms of equatorial differential 
rotation $\Omega_{\mathrm{eq}}$, $r_c=0.71$ and $d_c=0.025$. 
As usual, the dynamo equations can be made dimensionless by introducing the dynamo numbers
$C_\Omega = {R_\odot^2 \, \Omega_{\rm eq}}/{\etaT}$, 
$C_\alpha = {R_\odot\alpha_0}/{\etaT}$, 
$C_\omega = {R_\odot \omega}/{\etaT}$,
$C_u={R_\odot U}/{\etaT}$, 
where  $\omega$ is the frequency of the dynamo wave, and $U=u_\theta(r=R_\odot, \theta=45^\circ)$.
This linear dynamo problem  is solved with a finite-difference scheme for the radial dependence and a polynomial expansion for the 
angular dependence by imposing potential field as a boundary condition for the surface, and perfect conductor for the inner boundary.
Further details of the numerical approach  can be found in \cite{bo02,bo06} and also in \cite{jouve08}.

A reference solution can be then obtained in the region of high $C_u$, and its main properties are the following: 
$C_\alpha=2$, $C_\Omega=4.5 \times 10^5$, and $C_u=400$ which implies a period of about 24 years, a turbulent diffusivity 
$\etat=3.1 \times 10^{11} $  $\mathrm{cm}^2\; \mathrm{s}^{-1}$ and a (poleward) surface flow of about 18 $\mathrm{m\; s^{-1}}$.
\begin{figure}
\centering
\includegraphics[width=1\textwidth]{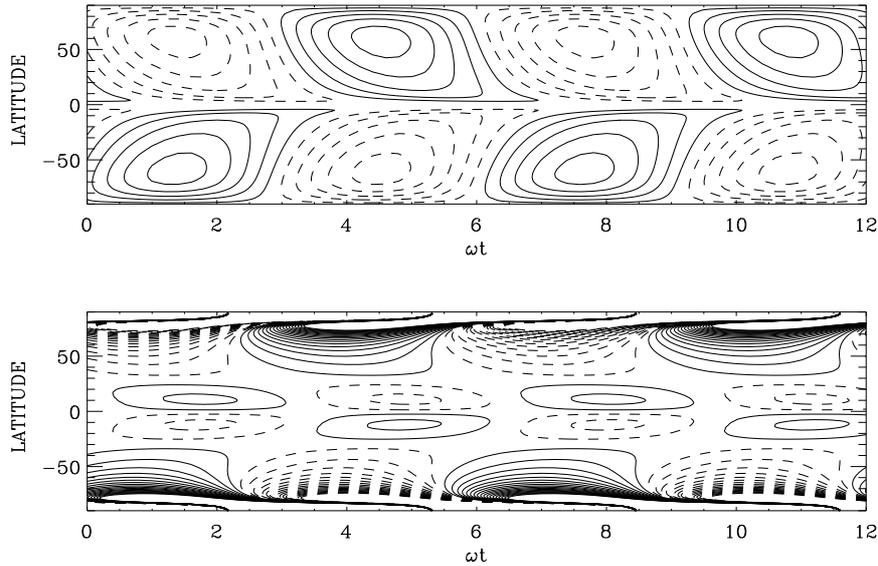} 
\label{fig:qq}
\caption{
Butterfly diagram for a solution  with meridional circulation included.
The isocontours for the toroidal field (upper panel) at the base of the convection zone
and the radial field on the top boundary are shown. Full and dashed lines show the positive and negative levels respectively.
The solution is obtained with
$C_\alpha=2$, $C_\Omega=4.5 \times 10^5$ and $C_u=400$ which implies a period of about 24 years a turbulent diffusivity 
$\etat=3.1 \times 10^{11}$  $\mathrm{cm}^2\; \mathrm{s}^{-1}$ and a (poleward) surface flow of about 18 $\mathrm{m\; s^{-1}}$.
It is interesting to notice that at low latitude the phase relation
is consistent with the observations, namely $B_r B_\phi <0$. 
This type of solution  also shows a strong preference for dipolar modes. In particular the critical solution for 
the symmetric mode has $C_\alpha = 3.70$, significantly greater than the antisymmetric mode. An essential ingredient
in order to get this property correct is the presence of a strong meridional circulation at the bottom of the convective zone. In particular in this case
the bottom flow is about 8.5 $\mathrm{m\; s^{-1}}$.\label{3tre}}
\end{figure}
The basic properties of this solution are depicted in Figure 3 where the butterfly diagram for the toroidal field at the bottom
of the convection zone, and of the radial field at the surface are shown. It is interesting to notice that at low latitude the phase relation
is consistent with the observations, namely $B_r B_\phi <0$. The other property of  this 
solution is that it shows a strong preference for dipolar modes. In particular the critical solution for 
the symmetric mode has $C_\alpha = 3.70$, significantly greater than the antisymmetric mode.
On the contrary, had we considered the case of an $\alpha$-effect uniformly distributed throughout the whole of the convection zone, 
the result would have been $C_\alpha=2.1$ for both symmetric and antisymmetric mode with a difference of about $1\%$
most probably due to numerics. The conclusion is that 
the parity of the solution is more likely to be a dipole if the $\alpha$-effect is located at the bottom of the convection zone.
{In our simulations the precise value of the ratio $\eta_{\rm c}/\etat$ was not crucial as we obtained basically the dynamo
solution for $\eta_{\rm c}/\etat=0.005$ and  $\eta_{\rm c}/\etat=0.05$ although in all of the simulations presented in this 
work the value  $\eta_{\rm c}/\etat=0.01$ has been chosen. 
For instance the period obtained for $\eta_{\rm c}/\etat=0.005$ was $24.9$ years, and for $\eta_{\rm c}/\etat=0.05$ was $24.5$ years.
This is not surprising as the penetration of the meridional 
flow is very weak as showed in the right panel of Figure (\ref{fig:stream}) and therefore the precise 
value of $\eta_c$ cannot significantly change the type of dynamo action.
In addition, changing the width of the $\alpha$ effect and of the turbulent
diffusivity also did not lead to significant changes in the solution, although we expect that the spatial extension of the turbulent
layer should be of the order of the tachocline width $d_\eta \approx 0.05$ solar radii.}

However, the serious, unsatisfactory aspect of advection dominated dynamo is the fact that the strength of the return flow is largely unknown, 
and the eddy diffusivity is about one order of magnitude greater than  would be expected on the basis of the standard 
mixing length theory. 

\section{Solar Dynamo from Subsurface Shear Instabilities}
\begin{figure}
\centering
\includegraphics[width=0.55\textwidth]{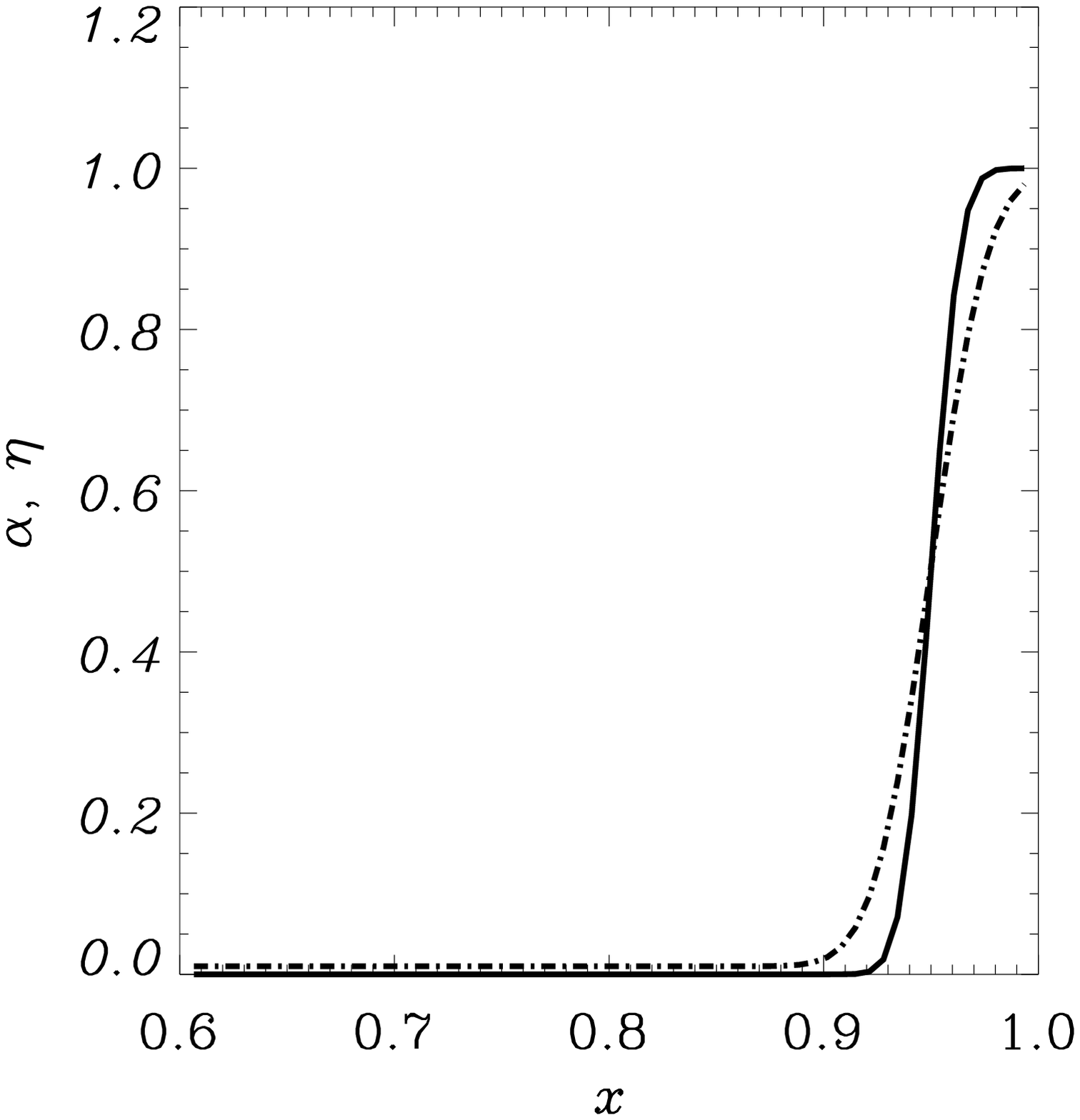}
\includegraphics[width=4cm, height=7cm]{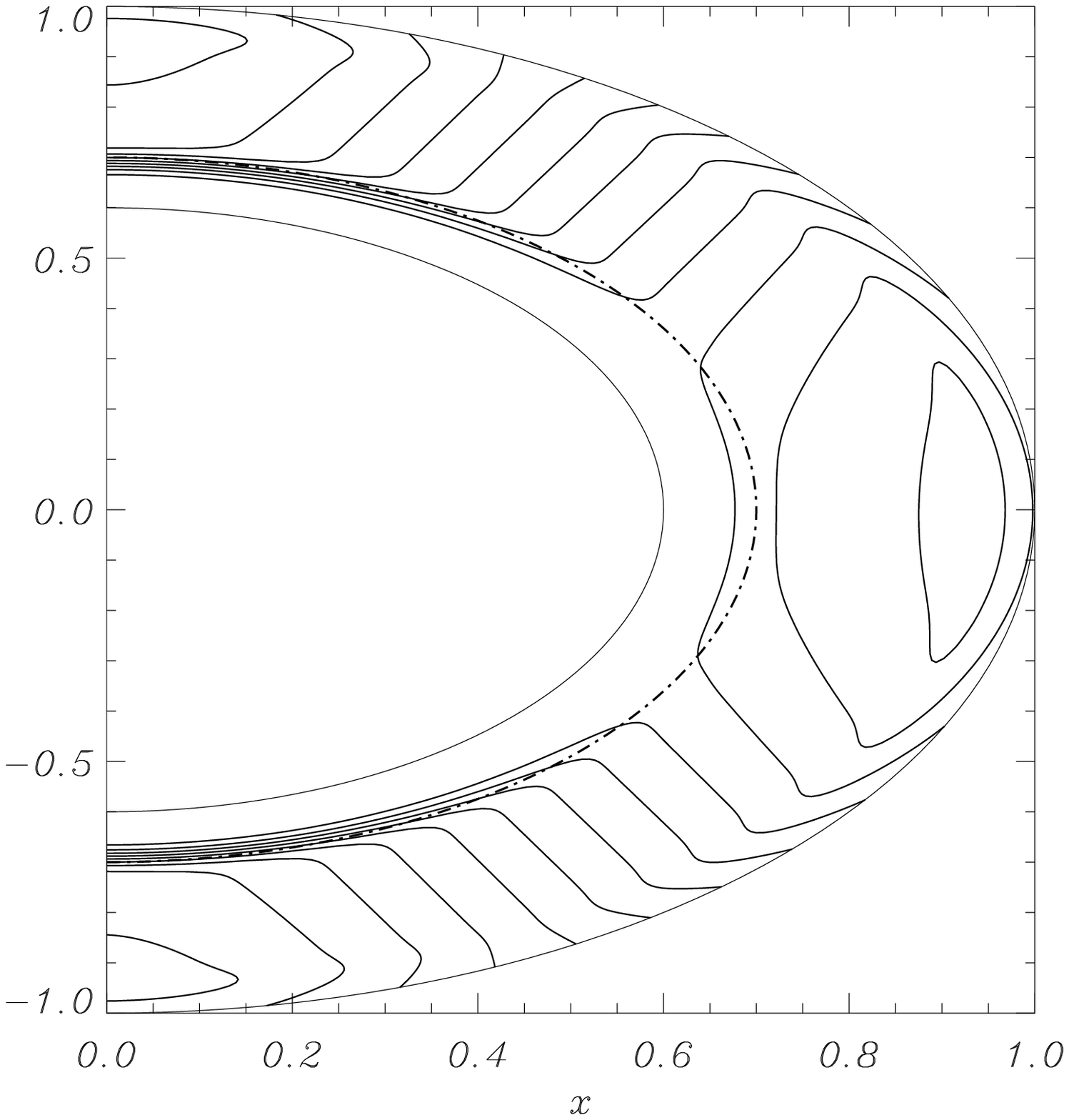}
\caption{Left panel: the $\alpha$-effect profile (solid line) and turbulent diffusivity (dot-dashed line) 
used for an $\alpha\Omega$ model with a dynamo action produced from subsurface shear instabilities.
Right panel: the isocontour lines of the differential rotation [$\Omega(r,\theta)$] 
in units of the equatorial rotation. Note the negative shear near the surface around $r/R_\odot \approx 0.95$.  }
\label{diro}
\end{figure}

In recent years the possibility that the dynamo operates in the subphotospheric layers of the Sun mostly 
driven by the negative gradient of the angular velocity near the surface has been proposed \citep{axel05}. 
It is interesting to see if a mean-field dynamo model can describe this case with the help of a more refined 
differential-rotation profile including the negative shear in the subsurface layers.
In order to discuss this issue a slightly modified version of the analytical approximation presented
in \cite{diktom} has been used. 
The radial profile of the $\alpha$-effect and turbulent diffusivity are depicted in Figure (\ref{diro}).
In particular the rotation rate is taken constant in the radiative interior $[\Omega_c]$ and the tachocline is located at the same
radius as in the model described in the previous session. 
It is assumed that has constant width $\approx \,0.05 R_\odot$, and at the top of the tachocline its rotation 
rate is given by  
\EQ
\Omega(r_{\rm cz},\theta) = \Omega_{\rm cz}+a_2 \cos^2 \theta +a_4 \cos^4 \theta
\EN
where $\Omega_{\rm cz}=-a_2/5- 3\; a_4/35$, $a_2=-61\; $nHz and $a_4=-73.5\; $nHz.  It is thus assumed that there
is a known negative gradient below the surface 
down to a radius $r_{s}=0.95 R_\odot$ and the latitudinal dependence of this shear layer is modeled by 
$P(\theta) = (p_0 + p_4 \cos^4 \theta\label{beta})/R_\odot$, 
so that 
\begin{eqnarray}
&&\Omega(r,\theta)=\Omega_c + \delta \left (\frac{r-r_c}{d_c}\right) Q(\theta) (r-r_{cz})+ \delta \left (\frac{r-r_s}{d_s}\right)[\Omega(r_{\rm cz},\theta)-\Omega_{\rm c}]\nonumber\\
&&+ \delta \left (\frac{r-r_s}{d_s}\right)[\Omega_{\rm eq}-\Omega_{\rm cz}-P(\theta)(r-R_\odot)-Q(\theta)(r-r_{\rm cz})]
\end{eqnarray}
and $Q(\theta) = (\Omega_{\rm eq}-\Omega_{\rm cz}+P(\theta)(R_\odot-r_{\rm s}))/(r-r_{\rm cz})$. For actual calculations the values $p_0=437\;$nHz and 
$p_4=-722\;$nHz have been chosen. In order to confine the magnetic field in the subsuface layers 
it is assumed that $\etat$ is also maximum near the surface, and it sharply decreases by about two orders of 
magnitude just below the supergranulation layer where the $\alpha$-effect is located.
\begin{figure}
\centering
\includegraphics[width=1\textwidth]{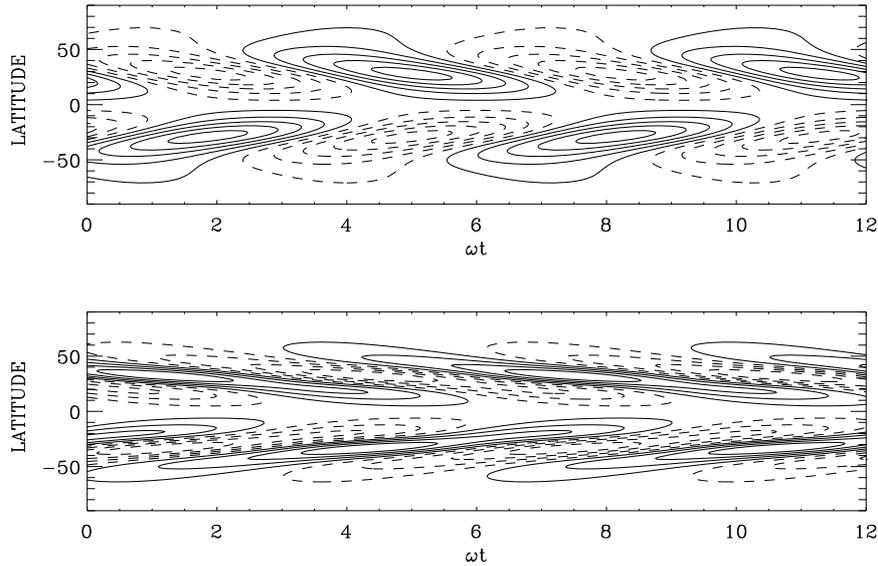} 
\label{zina}
\caption{
Butterfly diagram for a model with strong radial subsurface shear.
The isocontours for the toroidal field (upper panel) at $r=0.95 R$ and the radial field on the top boundary are shown. 
Full and dashed lines show the positive and negative levels respectively.
The solution is obtained with
$C_\alpha=18$ and $C_\Omega=1 \times 10^3$ which implies a period of about four years and a turbulent diffusivity 
$\etat=1.4 \times 10^{13} $  \cm$^2$s$^{-1}$.}
\end{figure}
The basic features of the (critical) dynamo solution in this case is depicted in Figure \ref{zina} for a solution with 
$C_\alpha=18$, $C_\Omega=1 \times 10^3$ which implies a period of about four years and a turbulent diffusivity 
$\etat=1.4 \times 10^{13} $  \cm$^2$s$^{-1}$. The period is clearly too small, but this is not surprising as the dynamo wave is basically entirely confined
in the surface layers as  can be seen in Figure (\ref{ribo}). 
The only possibility to match the solar period is to further increase the turbulent diffusivity in the supergranulation layer, but this would
imply that the spatial extension of the dynamo wave propagates deeply within the convective zone, 
thus preventing the negative radial shear to produce the correct
butterfly diagram. Moreover the parity of the solution is clearly symmetric because $C_\alpha=15$ for quadrupolar modes.   
\begin{figure}
\centering
\includegraphics[width=0.8\textwidth]{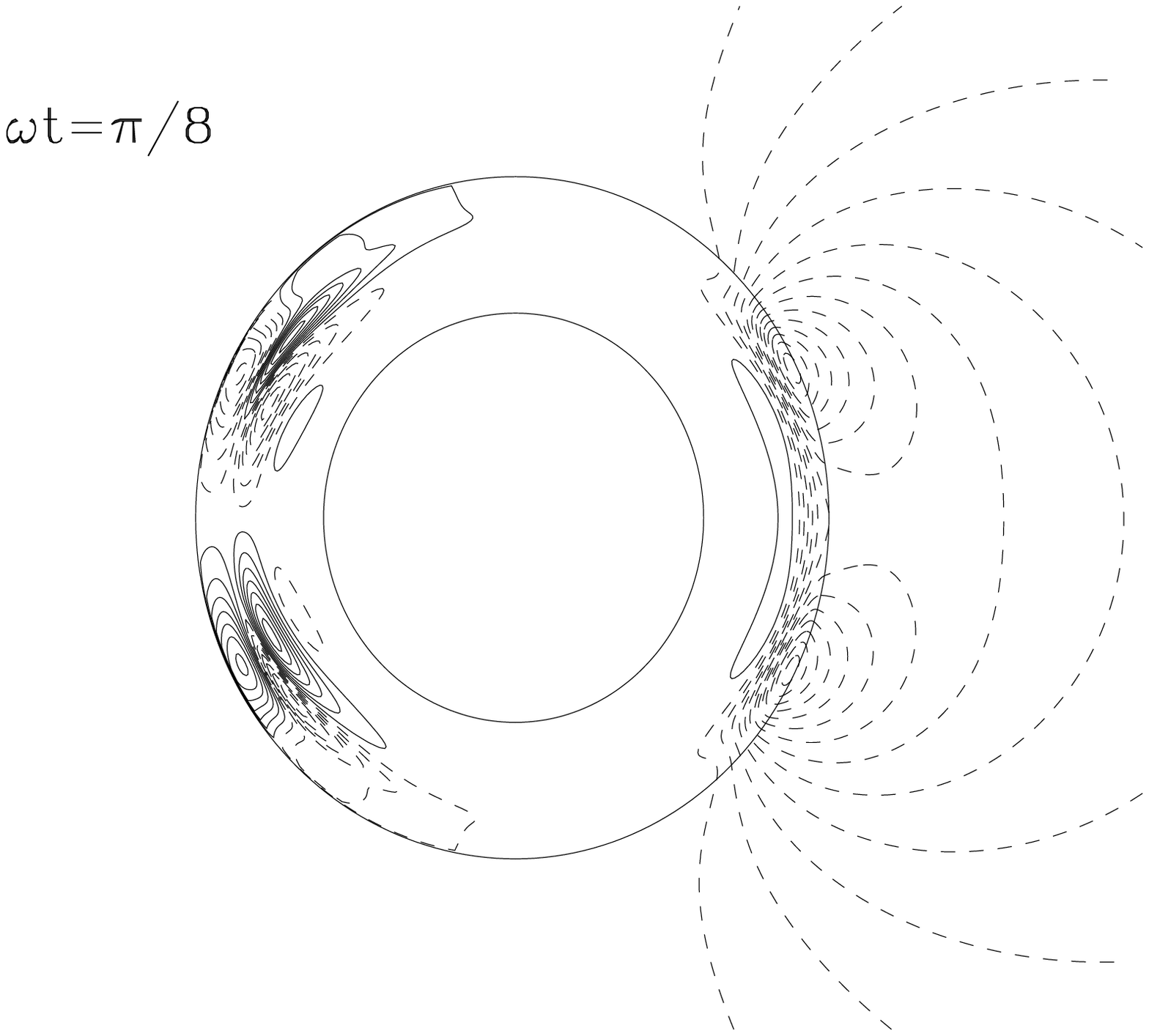}
\caption{
Toroidal and poloidal field configuration 
for a typical dynamo solution with a strong subsurface shear, 
and $C_\alpha=18$, $C_\Omega=1 \times 10^3$.
The left part are the isocontours line of the toroidal field, 
with solid line for negative $B_\phi$ and dashed line for positive value of the field. The right part represents of the streamlines of the poloidal field
given by contours of $A r \sin \theta$. Solid line are for negative values of $A$. Note that the dynamo action is confined around the region of strongest 
radial shear,  and, for this reason, the period tends to be rather short in general (less than ten years in these type of models).}
\label{ribo}
\end{figure}

\section{Conclusions}
Where is the $\alpha$-effect located in the Sun? Despite the difficulties present in models of flux-dominated dynamo (too low eddy diffusivity, unknown
strength and location of the return flow), mean-field models with an $\alpha$-effect located at the bottom of the convective zone are successful in 
reproducing several aspects of the solar activity cycle. In this model the origin of the $\alpha$-effect is due to the turbulent magnetic helicity and current helicity 
generated by the quasi-interchange instabilities below the tachocline and does not follow from mixing-length theory.
The most striking result comes from the parity of the solution: only with an $\alpha$-effect
located at the bottom of the convective zone the dipolar modes are most easily excited, at least at the linear level. This result also 
apply for models with the $\alpha$-effect generated by the subsurface shear, the solutions are mostly symmetric, 
rather than anti-symmetric. It would be nice to see if a more realistic surface boundary condition can solve this problem as proposed by \cite{vale}.

Although the models discussed in the previous session are kinematic, it is difficult to believe that the discussion of the previous session will drastically change  if a non-linearity via the $\alpha$-quenching is included in the models.  \cite{jouve08} present a code comparison between kinematic models and 
non-linear  models  that shows that the critical solutions are basically the same within the tolerance of different numerical schemes. 
On the other hand the advantage of the kinematic approach is to provide a  complete view of the spectrum of the dynamo waves that can be excited for a given set of dynamo numbers. This is clearly an important piece of information for the non-linear simulations, for instance for
investigating degenerate configurations due to the degeneracy in the spectrum.

An important constraint on the viability of a solar dynamo with a tachocline $\alpha$-effect is the fact that successful models must have a {\it positive} 
$\alpha$-effect: no migration is present for a negative $\alpha$-effect. 
It would be an important check for the theory to show that this is actually the case by means of 3D global numerical simulations
of the kink and quasi-interchange instability in spherical symmetry. Recent investigations in this direction in cylindrical symmetry have 
questioned this possibility \citep{gellert11} although further numerical and analytical work is needed before a firm conclusion can be reached.

\begin{acknowledgements}
I am grateful to Axel Brandenburg and Fabio Del Sordo for stimulating discussions.
I would also like to thank the organizers of the SDO meeting (in particular Sasha Kosovichev) and ESA for financial support. 
\end{acknowledgements}
%
%
\newcommand{\yastroph}[2]{ #1, astro-ph/#2}
\newcommand{\ycsf}[3]{ #1, {Chaos, Solitons \& Fractals,} {#2}, #3}
\newcommand{\yepl}[3]{ #1, {Europhys.\ Lett.,} {#2}, #3}
\newcommand{\yaj}[3]{ #1, {AJ,} {#2}, #3}
\newcommand{\yjgr}[3]{ #1, {J.\ Geophys.\ Res.,} {#2}, #3}
\newcommand{\ysol}[3]{ #1, {Sol.\ Phys.,} {#2}, #3}
\newcommand{\yapj}[3]{ #1, {\em Astrophys. J.}  {#2}, #3}
\newcommand{\ypasp}[3]{ #1, {PASP,} {#2}, #3}
\newcommand{\yapjl}[3]{ #1, {\em Astrophys. J.} {#2}, #3}
\newcommand{\yapjs}[3]{ #1, {ApJS,} {#2}, #3}
\newcommand{\yija}[3]{ #1, {Int.\ J.\ Astrobiol.,} {#2}, #3}
\newcommand{\yan}[3]{ #1, {Astron.\ Nachr.,} {#2}, #3}
\newcommand{\yzfa}[3]{ #1, {\em Z. Naturforsch.} {#2}, #3}
\newcommand{\ymhdn}[3]{ #1, {Magnetohydrodyn.} {#2}, #3}
\newcommand{\yana}[3]{ #1, {{\em Astron. Astrophys.}} {#2}, #3}
\newcommand{\yanas}[3]{ #1, {A\&AS,} {#2}, #3}
\newcommand{\yanar}[3]{ #1, {A\&A Rev.,} {#2}, #3}
\newcommand{\yass}[3]{ #1, {Ap\&SS,} {#2}, #3}
\newcommand{\sgafd}[1]{ #1, {Geophys.\ Astrophys.\ Fluid Dyn.,} submitted}
\newcommand{\ygafd}[3]{ #1, {Geophys.\ Astrophys.\ Fluid Dyn.,} {#2}, #3}
\newcommand{\ygrl}[3]{ #1, {Geophys.\ Res.\ Lett.,} {#2}, #3}
\newcommand{\ypasj}[3]{ #1, {Publ.\ Astron.\ Soc.\ Japan,} {#2}, #3}
\newcommand{\yjfm}[3]{ #1, {J.\ Fluid Mech.,} {#2}, #3}
\newcommand{\ypepi}[3]{ #1, {Phys.\ Earth Planet.\ Int.,} {#2}, #3}
\newcommand{\ypf}[3]{ #1, {Phys.\ Fluids,} {#2}, #3}
\newcommand{\ypfb}[3]{ #1, {Phys.\ Fluids B,} {#2}, #3}
\newcommand{\ypp}[3]{ #1, {Phys.\ Plasmas,} {#2}, #3}
\newcommand{\ysov}[3]{ #1, {Sov.\ Astron.,} {#2}, #3}
\newcommand{\ysovl}[3]{ #1, {Sov.\ Astron.\ Lett.,} {#2}, #3}
\newcommand{\yjetp}[3]{ #1, {Sov.\ Phys.\ JETP,} {#2}, #3}
\newcommand{\yphy}[3]{ #1, {Physica,} {#2}, #3}
\newcommand{\yaraa}[3]{ #1, {ARA\&A,} {#2}, #3}
\newcommand{\yanf}[3]{ #1, {Ann. Rev. Fluid Mech.,} {#2}, #3}
\newcommand{\yrpp}[3]{ #1, {Rep.\ Prog.\ Phys.,} {#2}, #3}
\newcommand{\yprs}[3]{ #1, {{\it Proc.\ Roy.\ Soc.\ Lond.},} {#2}, #3}
\newcommand{\yprt}[3]{ #1, {Phys.\ Rep.,} {#2}, #3}
\newcommand{\yprl}[3]{ #1, {Phys.\ Rev.\ Lett.,} {#2}, #3}
\newcommand{\yphl}[3]{ #1, {Phys.\ Lett.,} {#2}, #3}
\newcommand{\yptrs}[3]{ #1, {Phil.\ Trans.\ Roy.\ Soc.,} {#2}, #3}
\newcommand{\ymn}[3]{ #1, {\it Mon. Not. Roy. Astron. Soc.} {#2}, #3}
\newcommand{\ynat}[3]{ #1, {Nature,} {#2}, #3}
\newcommand{\yptrsa}[3]{ #1, {Phil. Trans. Roy. Soc. London A,} {#2}, #3}
\newcommand{\ysci}[3]{ #1, {Science,} {#2}, #3}
\newcommand{\ysph}[3]{ #1, {Solar Phys.,} {#2}, #3}
\newcommand{\ypr}[3]{ #1, {Phys.\ Rev.,} {#2}, #3}
\newcommand{\ypre}[3]{ #1, {Phys.\ Rev.\ E,} {#2}, #3}
\newcommand{\ypnas}[3]{ #1, {Proc.\ Nat.\ Acad.\ Sci.,} {#2}, #3}
\newcommand{\yicarus}[3]{ #1, {Icarus,} {#2}, #3}
\newcommand{\yspd}[3]{ #1, {Sov.\ Phys.\ Dokl.,} {#2}, #3}
\newcommand{\yjcp}[3]{ #1, {J.\ Comput.\ Phys.,} {#2}, #3}
\newcommand{\yjour}[4]{ #1, {#2}, {#3}, #4}
\newcommand{\yprep}[2]{ #1, {\sf #2}}
\newcommand{\ybook}[3]{ #1, {#2} (#3)}
\newcommand{\yproc}[5]{ #1, in {#3}, ed.\ #4 (#5), #2}
\newcommand{\pproc}[4]{ #1, in {#2}, ed.\ #3 (#4), (in press)}
\newcommand{\pprocc}[5]{ #1, in {#2}, ed.\ #3 (#4, #5)}
\newcommand{\pmn}[1]{ #1, {MNRAS}, to be published}
\newcommand{\pana}[1]{ #1, {A\&A}, to be published}
\newcommand{\papj}[1]{ #1, {ApJ}, to be published}
\newcommand{\ppapj}[3]{ #1, {ApJ}, {#2}, to be published in the #3 issue}
\newcommand{\sprl}[1]{ #1, {PRL}, submitted}
\newcommand{\sapj}[1]{ #1, {ApJ}, submitted}
\newcommand{\sana}[1]{ #1, {A\&A}, submitted}
\newcommand{\smn}[1]{ #1, {MNRAS}, submitted}

\end{article}
\end{document}